\documentclass[fleqn,usenatbib]{mnras}

\usepackage{newtxtext,newtxmath}

\usepackage[T1]{fontenc}
\usepackage{ae,aecompl}

\usepackage{graphicx}
\usepackage{amsmath}
\usepackage{amssymb}
\usepackage{subdepth}
\usepackage{multicol}
\usepackage{natbib}
\usepackage{booktabs,caption}
\usepackage[flushleft]{threeparttable}
\usepackage[normalem]{ulem}

\newcommand{\pcm}{\,cm$^{-3}$}
\newcommand*{\R}{\textcolor{red}}

\title[Nonthermal afterglow of GW170817]{Nonthermal afterglow of the binary neutron star merger GW170817: a more natural modeling of electron energy distribution leads to a qualitatively different new solution}

\author[H. Lin et al.]{Haoxiang Lin,$^{1}$\thanks{E-mail: haoxiang@astron.s.u-tokyo.ac.jp} Tomonori Totani,$^{1,2}$ Kenta Kiuchi$^{3}$ \\
$^{1}$Department of Astronomy, School of Science, The University of Tokyo, Tokyo 113-0033, Japan\\
$^{2}$Research Center for the Early Universe, School of Science, The University of Tokyo, Tokyo 113-0033, Japan\\
$^{3}$Center for Gravitational Physics, Yukawa Institute for Theoretical Physics, Kyoto University, Kyoto 606-8502, Japan
}

\date{Accepted XXX. Received YYY; in original form ZZZ}

\pubyear{2018}

\usepackage{etoolbox}
\makeatletter
\patchcmd\@combinedblfloats{\box\@outputbox}{\unvbox\@outputbox}{}{%
   \errmessage{\noexpand\@combinedblfloats could not be patched}%
}%
\makeatother

\begin{document}
\label{firstpage}
\pagerange{\pageref{firstpage}--\pageref{lastpage}}
\maketitle

\begin{abstract}

The observed nonthermal afterglow spectrum of the binary neutron star (BNS) merger GW170817 from radio to X-ray are consistent with synchrotron radiation by shock-accelerated electrons. However, previous afterglow modeling studies were based on a simplified assumption that the acceleration efficiency is extremely high, i.e. all electrons in the shock are accelerated as a nonthermal population. This affects the estimate of the minimum electron energy and hence $\nu_m$, the peak frequency of the afterglow spectrum. Here we present Bayesian fitting to the observed data with a more natural electron energy distribution, in which the acceleration efficiency is a free parameter. Interestingly, the maximum likelihood solutions are found with radio flux below $\nu_m$ in the early phase, in contrast to previous studies that found the radio frequency always above $\nu_m$. Therefore the $\nu_m$ passage through the radio band could have been clearly detected for GW170817, if sufficient low-frequency radio data had been taken in early time. In the new solutions, the lowest energy of electrons is found close to equipartition with the post shock protons, but only a small fraction ($<$10\%) of electrons are accelerated as nonthermal particles. The jet energy and interstellar medium density are increased by 1--2 orders of magnitude from the conventional modeling, though these are still consistent with other constraints. We encourage to take densely sampled low-frequency radio data in the early phase for future BNS merger events, which would potentially detect $\nu_m$ passage and give a strong constraint on electron energy distribution and particle acceleration efficiency. 

\end{abstract}


\section{Introduction} \label{sec:intro}
GW170817 is the first ever discovery of gravitational waves from a binary neutron star (BNS) merger event detected by advanced LIGO/VIRGO \citep{2017PhRvL.119p1101A}. This signal was soon followed by a campaign of broadband electromagnetic observations, leading to the detection of a gamma-ray burst (GRB) 170817A \citep{2017ApJ...848L..13A, 2017ApJ...848L..14G, 2017ApJ...848L..15S} and a kilonova SSS17a/AT2017gfo of synthesized mass $\sim$ $0.02$--$0.06M_{\odot}$ moving with a bulk velocity $\sim$ $0.1$--$0.3c$ \citep[e.g.][]{2017Natur.551...64A, 2017ApJ...848L..19C, 2017Sci...358.1556C, 2017ApJ...848L..17C, 2017Sci...358.1570D, 2017Sci...358.1565E, 2017Natur.551...80K, 2017Sci...358.1583K, 2017Natur.551...67P, 2017ApJ...848L..15S, 2017Sci...358.1574S, 2017Natur.551...75S, 2017PASJ...69..102T, 2017ApJ...848L..27T, 2017ApJ...848L..24V, 2017ApJ...851L..21V}. The source is located in NGC 4993, an elliptical galaxy at a distance of $\sim 40$ Mpc \citep{2017Sci...358.1556C}.

X-ray and radio afterglow emissions from GW170817/GRB 170817A were later detected at 9 and 16 days after the merger, respectively \citep{2017Sci...358.1579H, 2017Natur.551...71T}. The continued X-ray to radio monitoring during the first 360 days reveals a brightening of $\propto {\R t_{\rm obs}}^{0.8}$ till a turnover at $\sim$ 150 days, with a single and constant power-law spectral energy distribution $F_\nu \propto \nu^{-0.6}$ which is consistent with synchrotron radiation \citep[e.g.][]{2018ApJ...863L..18A, 2018A&A...613L...1D, 2018ApJ...858L..15D, 2017ApJ...848L..25H, 2017Sci...358.1579H, 2018NatAs...2..751L, 2017ApJ...848L..20M, 2018ApJ...856L..18M, 2018Natur.554..207M, 2018ApJ...867...57R, 2017Natur.551...71T, 2018MNRAS.tmpL..60T, 2018arXiv180806617V}. This implies that the nonthermal afterglow emissions from X-ray to radio originate from a single common electron population.

The rising pattern of the broadband flux has seriously challenged the straight-forward interpretation with either a homogeneous jet with a uniform energy distribution in angle, or a single-velocity spherical shell of expanding ejecta, both of which would generally predict a much faster rising flux $F_\nu \propto {\R t_{\rm obs}}^{3}$ \citep{2018Natur.554..207M, 2018MNRAS.478..407N}. Successful models require a more complicated outflow structure and can be categorized into (i) a structured jet containing an ultra-relativistic core and lower velocity wings \citep[e.g.][]{2018PhRvL.120x1103L, 2018ApJ...856L..18M}, (ii) a radially-stratified (quasi-)spherical outflow, which could be a mildly relativistic ``cocoon'' produced by a successful/chocked jet \citep[e.g.][]{2018Natur.554..207M, 2018ApJ...867...18N}, or the fast tail of the dynamical merger ejecta \citep[e.g.][]{2018Natur.554..207M, 2018ApJ...867...95H} initially driven by the shock wave formed at the collision front of BNS \citep[e.g.][]{2013ApJ...773...78B, 2013PhRvD..87b4001H, 2014MNRAS.437L...6K, 2017PhRvD..96h4060K}.

In spite of the large difference in geometry, there are too many model parameters and it is not easy to distinguish the jet and spherical models based only on afterglow fluxes \citep[e.g.][]{2018ApJ...856L..18M}. However, it is possible to break the degeneracy by their different temporal decay rates of light curves \citep{2018MNRAS.tmpL..60T}, the time of cooling break through X-ray bands \citep{2018ApJ...856L..18M} and afterglow radio imaging and polarization \citep{2018MNRAS.478.4128G}. The recent detection of superluminal motion \citep{2018Natur.561..355M} and the size measurement of the compact radio source \citep{2018arXiv180800469G} has confirmed the presence of a relativistic jet and strongly favors the jet-dominated model for GW170817, though the magnetic field configuration in the jet model is strongly constrained by the upper limit on linear polarization \citep{2018ApJ...861L..10C}.

In the previous modelings of nonthermal afterglow emission from GW170817 \citep{2018A&A...613L...1D, 2018MNRAS.478.4128G, 2018ApJ...867...95H, 2018ApJ...856L..18M, 2018Natur.554..207M, 2018ApJ...867...18N, 2018ApJ...867...57R, 2018MNRAS.tmpL..60T}, electron energy spectrum was treated in the same way as that conventionally used in GRB afterglow models \citep[e.g.][]{1998ApJ...497L..17S}. This assumes that all electrons in the shocked shell are accelerated into nonthermal power-law distribution, i.e., the injection efficiency of acceleration $f=1$, where $f$ is the number fraction of electrons that are injected to the shock acceleration process. This is clearly an oversimplification, and in reality, it is conceivable that a substantial fraction of electrons remains as thermal particles, as normally observed in supernova remnants \citep[e.g.][]{2001ApJ...546.1149L, 2003ApJ...589..827B}. Since the total number of nonthermal electrons is fixed in the previous studies, the minimum Lorentz factor of electrons in the shock frame, $\gamma_m$, is essentially determined by another parameter $\epsilon_e$, the fraction of total electron energy in the shock. However, physically $\gamma_m$ should be determined by the degree of equipartition between protons and electrons, which is independent of acceleration efficiency.

This oversimplified treatment has been used in GRB afterglow studies most likely because of the paucity of light curve data points in a wide wavelength range. In such cases, $f$ is in degeneracy with other model parameters including the explosion energy $E$, ambient density $n$, and energy fractions $\epsilon_B$, $\epsilon_e$ carried by magnetic field and nonthermal electrons respectively \citep{2005ApJ...627..861E}. If parameters are changed as $E \to E/f,\,n \to n/f,\,\epsilon_B \to f\epsilon_B,\, $ and $\epsilon_e \to f\epsilon_e$, the fit to observed data is preserved regardless of $f$. However, many data points of afterglow light curves, especially in radio bands, are available for GW170817. Some physical parameters are largely different from those of GRB afterglows, especially the shock velocity (ultra-relativistic for GRBs while only mildly relativistic for the bulk of BNS merger ejecta). Therefore, it is important to re-examine the afterglow modeling of GW170817 in terms of a more natural electron energy distribution in the trans-relativistic regime of the shock.

Here we present a Markov-Chain Monte-Carlo (MCMC) analysis to find the best-fit parameters of the nonthermal afterglow of GW170817, treating both $f$ and $\gamma_m$ as free parameters. We consider both the framework of a structured jet with a Gaussian energy profile in angle as well as a (quasi-)spherical outflow with radially stratified velocity. Even though the latter scenario has already been ruled out by observations of this event, it is still meaningful to examine how parameters of the spherical model are changed by our new treatment, considering the possibility that a spherical outflow may be found in other BNS merger events in the future. We will discuss how the nature and physical parameters of the best-fit solution are changed from the conventional modeling with the constraint of $f=1$.

In addition, we make some corrections and improvements to formulations
as follows. (1) It is common to consider a power-law kinetic energy
distribution of ejecta in the radially stratified spherical model as
$E(>u) \propto u^{-k}$ in a range $u_{\min} < u < u_{\max}$, where $u
= \beta \Gamma$ is a proper velocity of ejecta, $\beta$ a velocity
normalized by the light speed, and $\Gamma$ the Lorentz factor,
respectively. However, discontinuity in the cumulative distribution
$E(>u)$ at $u = u_{\max}$ means that there is a $\delta$-function like
concentration of ejecta mass. In this work we adopt a more reasonable
distribution which is power-law in the differential distribution,
$dE/du \propto u^{-k-1}$. (2) In previous studies an
ultra-relativistic limit of $\gamma_m \gg 1$ was used, but we use a
formula that is valid also in non-relativistic regime. (3) We exactly
calculate the velocity of a colliding ejecta shell velocity into the
shock region at a given time in the radially stratified spherical
model, while in some previous studies an approximation was used
\citep[e.g.][]{2018A&A...613L...1D, 2018MNRAS.478.4128G,
  2018Natur.554..207M, 2018MNRAS.tmpL..60T}. 

The structure of this paper is as follows. In Section \ref{sec:th} we present formulations of our model, and then we perform a MCMC analysis to the reported X-ray/optical/radio data and show the new constraints on model parameters in Section \ref{sec:mcmc}. We discuss implications of our results in Section \ref{sec:imp}, followed by a summary in Section \ref{sec:con}.

\section{Modeling} \label{sec:th}

\subsection{The Gaussian structured jet}
We consider a Gaussian angular profile for the energy distribution of the structured jet, as
\begin{align} \label{dist_jet}
    E_{k,\,{\rm iso}}(\theta) &= E_{c,\,{\rm iso}} \exp{\left(-\frac{\theta^2}{2\theta_c^2}\right)} \\
    \Gamma_0(\theta)-1 &= (\Gamma_c-1) \exp{\left(-\frac{\theta^2}{2\theta_c^2}\right)} \ ,
\end{align}
where $E_{k,\,{\rm iso}}(\theta)$ and $\Gamma_0(\theta)$ are the initial isotropic-equivalent jet kinetic energy (i.e., $4 \pi dE_{\rm jet}/d\Omega$ where $dE_{\rm jet}/d\Omega$ is the jet energy per unit solid angle, and $E_{\rm jet} \approx E_{c,\,{\rm iso}} \pi \theta_c^2$) and the Lorentz factor towards the direction of angle $\theta$ from the jet axis, and $E_{c,\,{\rm iso}}$ and $\Gamma_c$ are the model parameters. This assumes that the jet contains a constant rest mass per unit solid angle, following previous studies \citep[e.g.][]{2018MNRAS.478.4128G, 2018arXiv180610596H}.

\subsection{The radially stratified spherical outflow} 
We parametrize the distribution of ejecta energy over its proper velocity $u$ as a power law profile, $dE_{\rm sph}/du \propto u^{-k-1}$ in the range of $u_{\rm min} < u < u_{\rm max}$, and then the integrated form becomes
\begin{equation} \label{dist_sph}
  E_{\rm sph}(>u) = E_{k,\,{\rm iso}} \frac{u^{-k} - u_{\rm max}^{-k}}{u_{\rm min}^{-k} - u_{\rm max}^{-k}} \ \ \ \ \ (u_{\rm min} < u < u_{\rm max}) \ ,
\end{equation}
where $E_{k,\,{\rm iso}}$ is the total kinetic energy carried by the outflow. Note that a power law profile in the integrated from, $E_{\rm sph}(>u) = E_{k,\,{\rm iso}} (u/u_{\rm min})^{-k}$, has been commonly assumed in the literature \citep[e.g.][]{2018A&A...613L...1D, 2018MNRAS.478.4128G, 2018Natur.554..207M, 2018MNRAS.tmpL..60T}, but this introduces a discontinuity at $u_{\max}$ and hence a non-vanishing amount of energy $E_{k,\,{\rm iso}}(u_{\rm max}/u_{\rm min})^{-k}$ is concentrating at $u_{\rm max}$ like a $\delta$-function without any physical motivation. On the other hand, the profile (\ref{dist_sph}) properly describes the fast tail as $E_{\rm sph}(>u)$ vanishes at $u=u_{\rm max}$, and still asymptotically follows $E(>u) \propto u^{-k}$ when $u < u_{\rm max}$. We note that, however, this correction does not result in significant change of the model fitting, because most of the energy is carried by lowest velocity material.

\subsection{Shock dynamics}

Fast outflow generates a shock wave propagating into the circumburst medium or interstellar medium (ISM), which is slowed down by gradually sweeping up the medium. In this work we consider only the case of a uniform ISM density, $n$. We use a simple model in which the dynamics of the shock propagating into one direction is treated as a closed box, ignoring interactions between flows into different directions. Then the dynamics of the shock wave into an angle $\theta$ is obtained by solving the following equations:
\begin{align}
  \label{Inj}
  & E_{\rm inj}(\theta) = M[R(\theta)] \, c^2u_s^2(\theta) \\
  & u_s(\theta) = \frac{\dot{R}(\theta)}{c} \; \left\{ 1 - \left[ \frac{\dot{R}(\theta)}{c} \right]^2 \right\}^{-\frac{1}{2}} \ ,
\end{align}
where $E_{\rm inj}$ is the isotropic-equivalent energy injected into the shock until a post-merger time $t$ (in the burster frame, i.e., the lab frame), $R(\theta)$ and $u_s(\theta)$ are the radius and proper velocity of the shock front to angle $\theta$, respectively, $M[R(\theta)] = 4\pi R^3(\theta) n m_p/3$ the isotropic-equivalent swept-up ISM mass up to radius $R(\theta)$, and $m_p$ the proton mass. In this work we consider only the case of adiabatic shock evolution, which is assumed in the above equations.

In the jet scenario only one single injection occurs with isotropic-equivalent initial energy $E_{k,\,{\rm iso}}(\theta)$, but an amount of energy $(\Gamma_s-1)m_0c^2$ ($\Gamma_s = \sqrt{u_s^2-1}$ is the shock Lorentz factor) is carried by the corresponding initial rest mass $m_0 = E_{k,\,{\rm iso}}/[(\Gamma_0-1)c^2]$ and does not contribute to acceleration of the shocked external medium, and thus should be excluded from the energy injection:
\begin{equation}
  E_{\rm inj}(\theta) = E_{k,\,{\rm iso}}(\theta) \left[ 1 - \frac{\Gamma_s(\theta)-1}{\Gamma_0(\theta)-1} \right] \ .
\end{equation}

In the stratified spherical outflow scenario the injection is continuous and independent of $\theta$:
\begin{equation}
  E_{\rm inj} = E_{\rm sph}(>u_{\rm col}) \ ,
\end{equation}
where $u_{\rm col}$ is the velocity of the freely-expanding ejecta that collides into the shocked shell at time $t$, and hence given by
$\beta_{\rm col} = R/(ct)$ and
\begin{equation}
  u_{\rm col} = \frac{R}{\sqrt{c^2t^2-R^2}} \ .
\end{equation}
In previous studies an approximation of $u_{\rm col} = u_s$ has been used \citep[e.g.][]{2013MNRAS.430.2121P}, but here we use this exact formula to determine the shock motion.

\subsection{Electron energy distribution} \label{sec:edist}

The synchrotron emission theory formulated in \cite{1998ApJ...497L..17S} assumes that (i) the nonthermal electrons follow a power-law energy distribution with a minimum Lorentz factor $\gamma_m$: $dN_e/d\gamma_e \propto \gamma_e^{-p}$, $\gamma_e \geq \gamma_m$, (ii) a fraction $\epsilon_e$ of shock thermal energy goes to the nonthermal electrons and (iii) all electrons in the swept-up matter, $N_e = f M(R)/m_p$, are accelerated as nonthermal particles. This means a rather unrealistic case of 100\% injection efficiency ($f=1$) of particle acceleration. Then $\gamma_m$ is simply related to $\epsilon_e$ under a fixed $p$ as:
\begin{equation} \label{Sari}
  \gamma_m = \epsilon_e \frac{p-2}{p-1} \frac{m_p}{m_e} \Gamma_s \ ,
\end{equation}
where $m_e$ is the electron mass. It should also be noted that this formulation assumes the ultra-relativistic limit of $\gamma_m \gg 1$ and $\Gamma_s \gg 1$, though some of previous studies on GW170817 used this without modification. We note that there is a factor of order unity difference between the shock Lorentz factor $\Gamma_s$ and that of the shocked matter \citep[e.g.][]{1976PhFl...19.1130B}, but we follow previous studies and ignore it in the calculations hereafter.

Here we utilize a new formulation of electron energy distribution so that it properly describes the two degrees of freedom about $f$ and $\gamma_m$, which is consistent throughout the trans-relativistic regime. First we note that $\gamma_m$ should physically be related to the degree of equipartition between non-accelerated protons and electrons. Consider a stream of cold ISM passing through a shock front. The ISM protons and electrons are scattered at the shock front to become isotropic, leading to a post-shock proton ``temperature'' $k_B T_p \sim m_p c^2 (\Gamma_s-1)$ and a colder post-shock electron ``temperature'' $k_B T_e \sim m_e c^2 (\Gamma_s-1)$, where $k_B$ is the Boltzmann constant. These electrons can be heated up to equipartition with the protons. It is therefore natural to consider that the minimum injection energy of electron acceleration $m_e c^2 \gamma_m$ is between $k_B T_e$ and $k_B T_p$. Here we introduce a model parameter $\eta_e$ as
\begin{equation} \label{gamma_m}
  \gamma_m = \eta_e\frac{m_p}{m_e}(\Gamma_s - 1) \ ,
\end{equation}
and hence $\eta_e \sim 1$ in the case electron-proton equipartition and $\eta_e \sim m_e/m_p$ in the case of no energy transfer from protons to electrons. We note that $\eta_e$ is generally found to be $0.1$--$1$ in particle-in-cell (PIC) simulations of particle shock acceleration \citep[e.g.][]{2011ApJ...726...75S, 2015PhRvL.114h5003P}.

Given $\gamma_m$ (or $\eta_e$), the energy fraction $\epsilon_e$ is related to the acceleration efficiency $f$ as
\begin{equation}
  m_e \langle\gamma_e\rangle = \frac{\epsilon_e}{f} m_p(\Gamma_s-1) \ ,
\end{equation}
where $\langle\gamma_e\rangle = \gamma_m(p-1)/(p-2)$ is the mean electron energy. In this paper we take $\eta_e$, $\epsilon_e$, and $p$ to be the model parameters to determine the electron energy distribution, and $f$ is expressed as a function of them as
\begin{equation} \label{f}
 f = \frac{\epsilon_e}{\eta_e}\frac{p-2}{p-1} \ .
\end{equation}
By putting this with $f=1$ into eq.(\ref{gamma_m}), One can find the correspondence to eq.(\ref{Sari}) of the conventional model \cite{1998ApJ...497L..17S} in which $\gamma_m$ is essentially determined by $\epsilon_e$. 

When the shock becomes non-relativistic (i.e. $\Gamma_s - 1 \ll 1$), e.g. in the late-time afterglow phase, eq.(\ref{gamma_m}) could lead to a $\gamma_m$ smaller than 1 and becomes invalid. Furthermore, even when $\gamma_m$ is sufficiently close to 1, the majority of electrons will emit cyclotron radiation at a single frequency instead, and the formulation for synchrotron radiation below is no longer applicable. Therefore, we manually set $\gamma_e = 2$ as a lower limit for those electrons that mainly contribute to the synchrotron emission, and replace $\gamma_m$ with $\max(\gamma_m, 2)$. Correspondingly, the number of ``synchrotron-emitting'' electrons becomes smaller by a factor of $\gamma_m^{p-1}/2^{p-1}$, and thus $N_e = f M(R)/m_p$ should be replaced simultaneously with $N_e = f M(R)/m_p \times \min(1, \gamma_m^{p-1}/2^{p-1})$.

\subsection{Synchrotron emission}

In the following primes indicate quantities measured in the shock comoving frame. The synchrotron flux emitted from the shock region is calculated following \cite{1998ApJ...497L..17S}, but the expressions for $\gamma_m$ and $f$ are replaced by eq.(\ref{gamma_m}) and (\ref{f}) derived in Section \ref{sec:edist}. Furthermore, the formula for magnetic field $B'$ is revised to make it applicable in trans-relativistic regime:
\begin{equation} \label{B}
  B' = \left[ 8\pi\epsilon_B \frac{\hat{\gamma}\Gamma_s+1}{\hat{\gamma}-1} n m_p c^2 (\Gamma_s-1) \right]^{1/2} .
\end{equation}
Here we used the shock jump condition given in \cite{1976PhFl...19.1130B}, and the adiabatic index of the shocked gas is modeled in the mono-energetic gas approximation to be $\hat{\gamma} = (4\Gamma_s+1)/(3\Gamma_s)$ so that $\hat{\gamma} = 4/3$ and $5/3$ respectively for relativistic and non-relativistic gas (\citealt{1971ApJ...165..147M}, see also \citealt{2011ApJ...733...86U}).

In the shock comoving frame, the synchrotron photon frequencies corresponding to $\gamma_m$ and the critical electron Lorentz factor affected by radiative cooling $\gamma_c = (6\pi m_e c \Gamma_s)/(\sigma_T B'^2t)$, where $\sigma_T$ is the Thomson cross section, are
\begin{equation}
  \nu_{m,\,c}' = \frac{1}{2\pi} \frac{e B'}{m_e c} \gamma_{m,\,c}^2 \ .
\end{equation}
The maximum synchrotron emitting power $P'_{\rm max}$ per unit frequency is given as
\begin{equation}
  P'_{\rm max} = \frac{8\pi}{9} \frac{e^3B'}{m_e c^2} N_e \ ,
\end{equation}
where we adopt the same constant factor $8\pi/9$ following the approximation in \cite{1998ApJ...497L..17S} for a direct comparison, while the exact numerical factor is $\approx 1.33$ \citep[e.g.][]{2002ApJ...568..820G}. 
When $\nu'_m < \nu'_c$,
\begin{eqnarray}
  P'_{\nu'}/P'_{\rm max} = \begin{cases}
          \left(\nu_c'/\nu_m'\right)^{-(p-1)/2}\left(\nu'/\nu_c'\right)^{-p/2}
             \qquad (\nu_c' < \nu') \\
          \left(\nu'/\nu_m'\right)^{-(p-1)/2} \qquad (\nu_m' < \nu' < \nu_c') \\
          \left(\nu'/\nu_m'\right)^{1/3} \qquad (\nu' < \nu_m')
            \end{cases}
\end{eqnarray}
and when $\nu'_m > \nu'_c$,
\begin{eqnarray}
  P'_{\nu'}/P'_{\rm max} = \begin{cases}
              \left(\nu_m'/\nu_c'\right)^{-1/2}\left(\nu'/\nu_m'\right)^{-p/2}
                \qquad (\nu_m' < \nu') \\
              \left(\nu'/\nu_c'\right)^{-1/2} \qquad (\nu_c' < \nu' < \nu_m') \\
              \left(\nu'/\nu_c'\right)^{1/3} \qquad (\nu' < \nu_c') \ .
            \end{cases}
\end{eqnarray}
For simplicity we ignore synchrotron self-absorption because it is not expected to influence the observed frequency range of GW170817.

The flux density $F_{\nu}$ eventually received by the observer at luminosity distance $D_L$, time ${\R t_{\rm obs}}$ and an observing frequency $\nu$ is given by:
\begin{equation}
  F_{\nu}(\nu, {\R t_{\rm obs}}) = \frac{1+z}{4\pi D_L^2} \int^{2\pi}_{0} \frac{d\varphi}{2\pi} \int^{\pi}_{0} \frac{\sin{\theta}d\theta}{2} \delta_D^3 \, P'_{\nu'}(\nu', R) \ ,
\end{equation}
where $\delta_D = [\Gamma_s(1-\beta_s\mu)]^{-1}$ is the Doppler factor, and $\mu$ is the cosine angle of the velocity of emitting matter from the direction to the observer. Without loss of generality, we take a spherical coordinate system in which the jet axis coincides with zero polar angle ($\theta=0$), and the observer's direction at $\theta_v$ and zero azimuthal angle ($\varphi_v=0$). Then $\mu$ is given by
\begin{equation}
    \mu = \cos{\theta}\cos{\theta_v} + \sin{\theta}\cos{\varphi}\sin{\theta_v}\ .
\end{equation}
Integration is done with $\nu'(\mu) = (1+z)\nu/\delta_D(\mu)$ and $R$ replaced as a function of ${\R t_{\rm obs}}$, $\theta$ and $\varphi$, i.e. the equal arrival time surface, by solving \citep[e.g.][]{1999ApJ...513..679G}
\begin{equation}
    t(R, \theta, \phi) - \frac{R}{c}\mu(\theta, \varphi) = \frac{{\R t_{\rm obs}}}{1+z} \ .
\end{equation}

\begin{table*}
  \caption{Constraints on the Gaussian structured jet model parameters, which are further divided into two categories: one fixing the electron acceleration efficiency $f=1$ and the other allowing $f$ to vary freely ($f$ free). We chose a uniform distribution for each prior, and the median values of one-dimensional posterior distributions of each parameter are presented with the symmetric 68\% uncertainties (i.e. the 16\% and 84\% quantiles). The bottom row also shows the calculated jet energy $E_{\rm jet} = E_{c,\,{\rm iso}}\pi\theta_c^2$, and the corresponding range of injection efficiency $f$, calculated by the posterior distribution of $\epsilon_e$ and $\eta_e$ via eq.(\ref{f}).
  }
  \label{tab:jet}
  \bgroup
  \def\arraystretch{1.5}%
  \begin{center}
    \begin{tabular}{cccccccc}
      \toprule
       & \multicolumn{4}{c}{Jet $f=1$} & & \multicolumn{2}{c}{Jet $f$ free} \\
      \cline{2-5} \cline{7-8} \\
      Parameter & 1D dist.$^a$ & best-fit$^b$ & 1D dist.$^a$ ($\nu_{\rm obs} > \nu_m$) & best-fit$^b$ ($\nu_{\rm obs} > \nu_m$) & & 1D dist.$^a$ & best-fit$^b$ \\
      \midrule
      $\log_{10} (E_{c,\,{\rm iso}}/{\rm erg})$ & $51.05^{+0.51}_{-0.37}$ & $51.19$ & $52.67^{+0.55}_{-0.62}$ & $52.33$ & & $52.38^{+0.93}_{-0.90}$ & $52.25$ \\
      $\theta_c$ & $0.08^{+0.03}_{-0.02}$ & $0.09$ & $0.07^{+0.02}_{-0.01}$ & $0.08$ & & $0.08^{+0.03}_{-0.02}$ & $0.10$ \\
      $\theta_v$ & $0.45^{+0.14}_{-0.09} $ & $0.47$ & $0.42^{+0.10}_{-0.06}$ & $0.50$ & & $0.44^{+0.14}_{-0.10}$ & $0.50$ \\
      $\log_{10}\Gamma_c$ & $3.06^{+0.30}_{-0.29}$ & $2.93$ & $3.58^{+0.31}_{-0.45}$ & $3.86$ & & $3.04^{+0.28}_{-0.29}$ & $2.97$ \\
      \midrule
      $\log_{10} (n/{\rm cm}^{-3})$ & $-3.71^{+0.61}_{-0.65}$ & $-3.57$ & $-2.36^{+0.71}_{-0.75}$ & $-2.18$ & & $-2.49^{+1.05}_{-1.08}$ & $-2.28$ \\
      $\log_{10}\epsilon_B$ & $-2.57^{+0.89}_{-1.01}$ & $-3.12$ & $-4.30^{+1.38}_{-1.14}$ & $-4.60$ & & $-4.13^{+1.41}_{-1.20}$ & $-4.76$ \\
      $\log_{10}\epsilon_e$ & $-0.22^{+0.15}_{-0.21}$ & $-0.10$ & $-1.36^{+0.46}_{-0.57}$ & $-0.83$ & & $-1.28^{+0.81}_{-1.18}$ & $-0.80$ \\
      $p$ & $2.18^{+0.01}_{-0.01}$ & $2.19$ & $2.16^{+0.01}_{-0.01}$ & $2.16$ & & $2.18^{+0.01}_{-0.01}$ & $2.18$ \\
      $\log_{10}\eta_e$ & --- & --- & --- & --- & & $-1.02^{+0.23}_{-0.25}$ & $-0.80$ \\
      \midrule
      $\log_{10} E_{\rm jet}$ & $49.44^{+0.38}_{-0.30}$ & $49.46$ & $50.84^{+0.53}_{-0.60}$ & $50.63$ & & $50.74^{+0.89}_{-0.87}$ & $50.75$ \\
      $\log_{10} f$ & $0$ & $0$ & $0$ & $0$ & & $-1.09^{+0.75}_{-1.08}$ & $-0.83$ \\
      \midrule
      $\chi^2$ &  & $114$  &  & $140$ &  &  & $114$ \\
      \bottomrule
    \end{tabular}
  \end{center}
  \egroup
  \begin{tablenotes}
  \item $^a$Median with 1-$\sigma$ uncertainties ($16$\% and $84$\% quantiles) in one-dimensional probability distribution.
  \item $^b$The model parameters for the maximum of the posterior probability density function.
  \end{tablenotes}
\end{table*}

\begin{table*}
  \caption{Same as Table \ref{tab:jet} but with stratified spherical outflow model.}
  \label{tab:sph}
  \bgroup
  \def\arraystretch{1.5}%
  \begin{center}
    \begin{tabular}{cccccc}
      \toprule
      & \multicolumn{2}{c}{Sph $f=1$} & & \multicolumn{2}{c}{Sph $f$ free} \\
      \cline{2-3} \cline{5-6} \\
      Parameter & 1D dist. & best-fit & & 1D dist. & best-fit \\
      \midrule
      $\log_{10} (E_{k,\,{\rm iso}}/{\rm erg})$ & $50.15^{+1.27}_{-0.63}$ & $50.58$ & & $51.60^{+1.65}_{-1.70}$ & $50.18$ \\
      $\log_{10} u_{\rm max}$ & $1.20^{+0.53}_{-0.45}$ & $0.55$ & & $1.31^{+0.46}_{-0.46}$ & $0.59$ \\
      $\log_{10} u_{\rm min}$ & $0.40^{+0.02}_{-0.08} $ & $0.24$ & & $0.34^{+0.07}_{-0.02}$ & $0.24$ \\
      $k$ & $5.66^{+0.55}_{-0.32}$ & $6.21$ & & $5.69^{+0.45}_{-0.40}$ & $5.79$ \\
      \midrule
      $\log_{10} (n/{\rm cm}^{-3})$ & $-3.68^{+1.21}_{-0.67}$ & $-2.25$ & & $-2.05^{+1.66}_{-1.71}$ & $-2.69$ \\
      $\log_{10}\epsilon_B$ & $-0.94^{+0.61}_{-1.21}$ & $-1.94$ & & $-2.57^{+1.71}_{-1.66}$ & $-1.62$ \\
      $\log_{10}\epsilon_e$ & $-1.74^{+0.57}_{-1.22}$ & $-2.14$ & & $-3.36^{+1.70}_{-1.65}$ & $-1.93$ \\
      $p$ & $2.15^{+0.01}_{-0.01}$ & $2.15$ & & $2.16^{+0.01}_{-0.01}$ & $2.17$ \\
      $\log_{10}\eta_e$ & --- & --- & & $-1.04^{+0.08}_{-0.12}$ & $-0.87$ \\
      \midrule
      $\log_{10} f$ & $0$ & $0$ & & $-3.17^{+1.71}_{-1.64}$ & $-1.90$ \\
      \midrule
      $\chi^2$ &  & $148$ & &  & $125$ \\
      \bottomrule
    \end{tabular}
  \end{center}
  \egroup
\end{table*}

\begin{figure*}
\begin{center}
\begin{multicols}{2}
    \includegraphics[width=1\columnwidth]{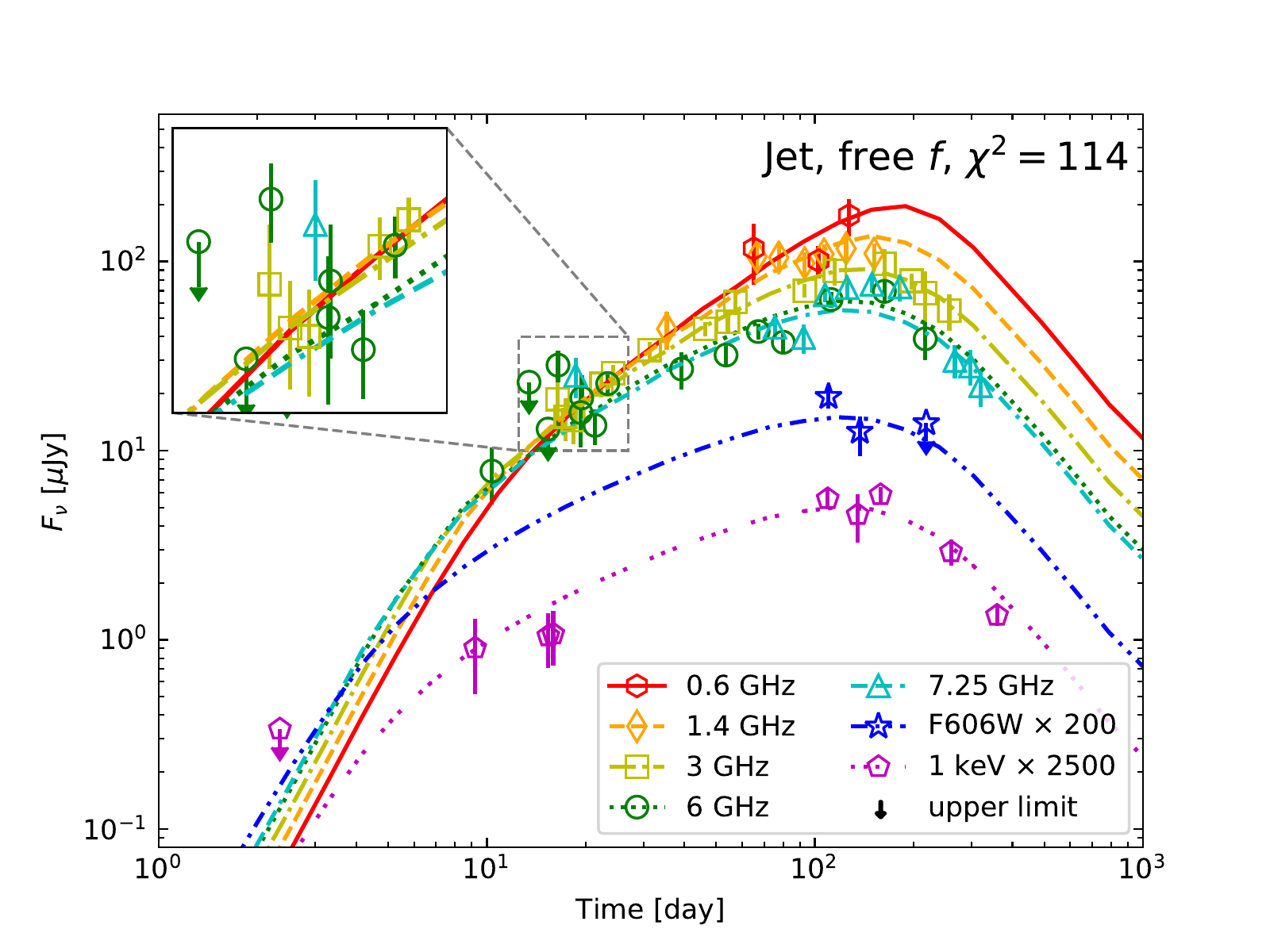}%\par
    \includegraphics[width=1\columnwidth]{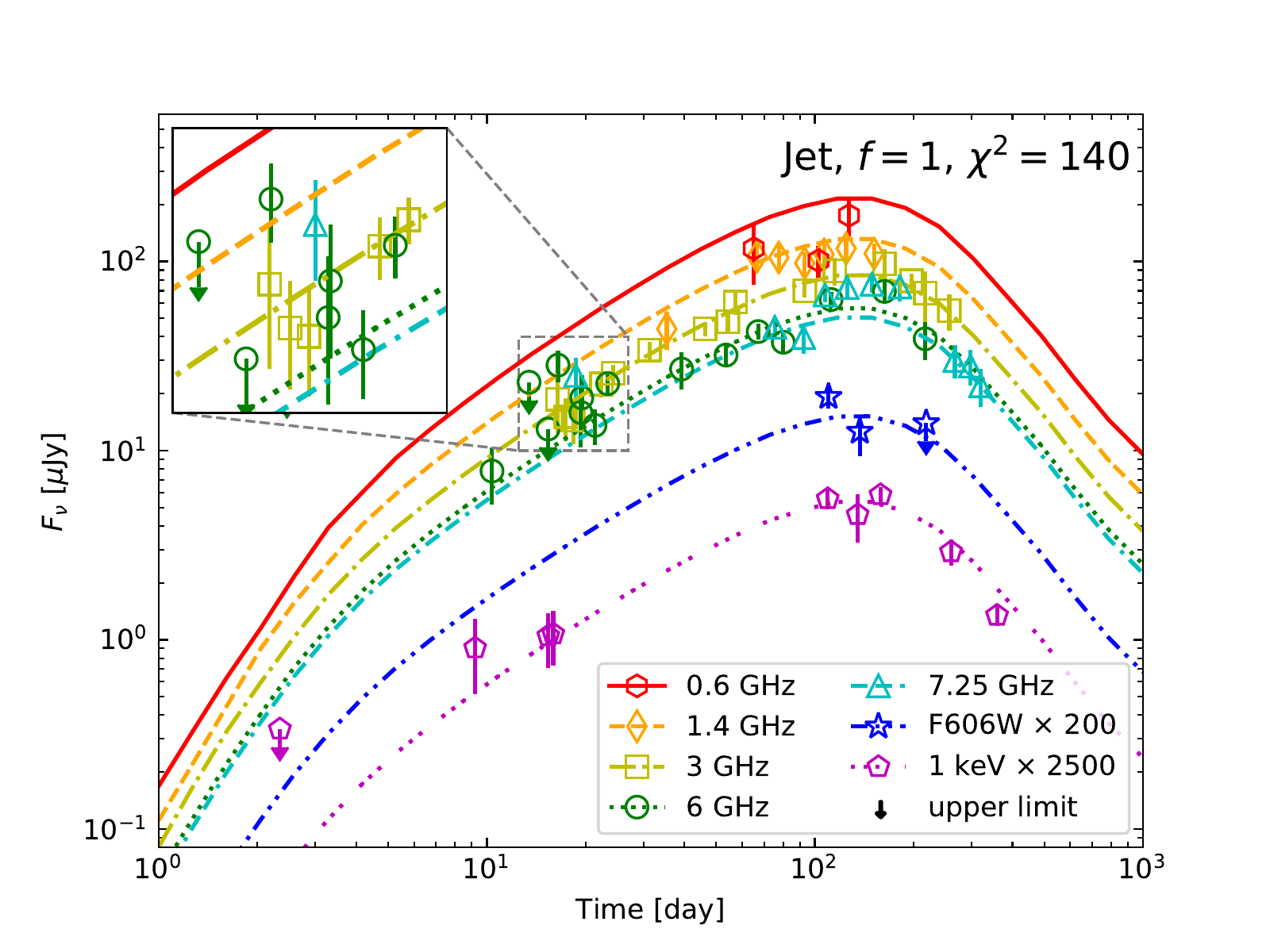}%\par
\end{multicols}
\begin{multicols}{2}
    \includegraphics[width=1\columnwidth]{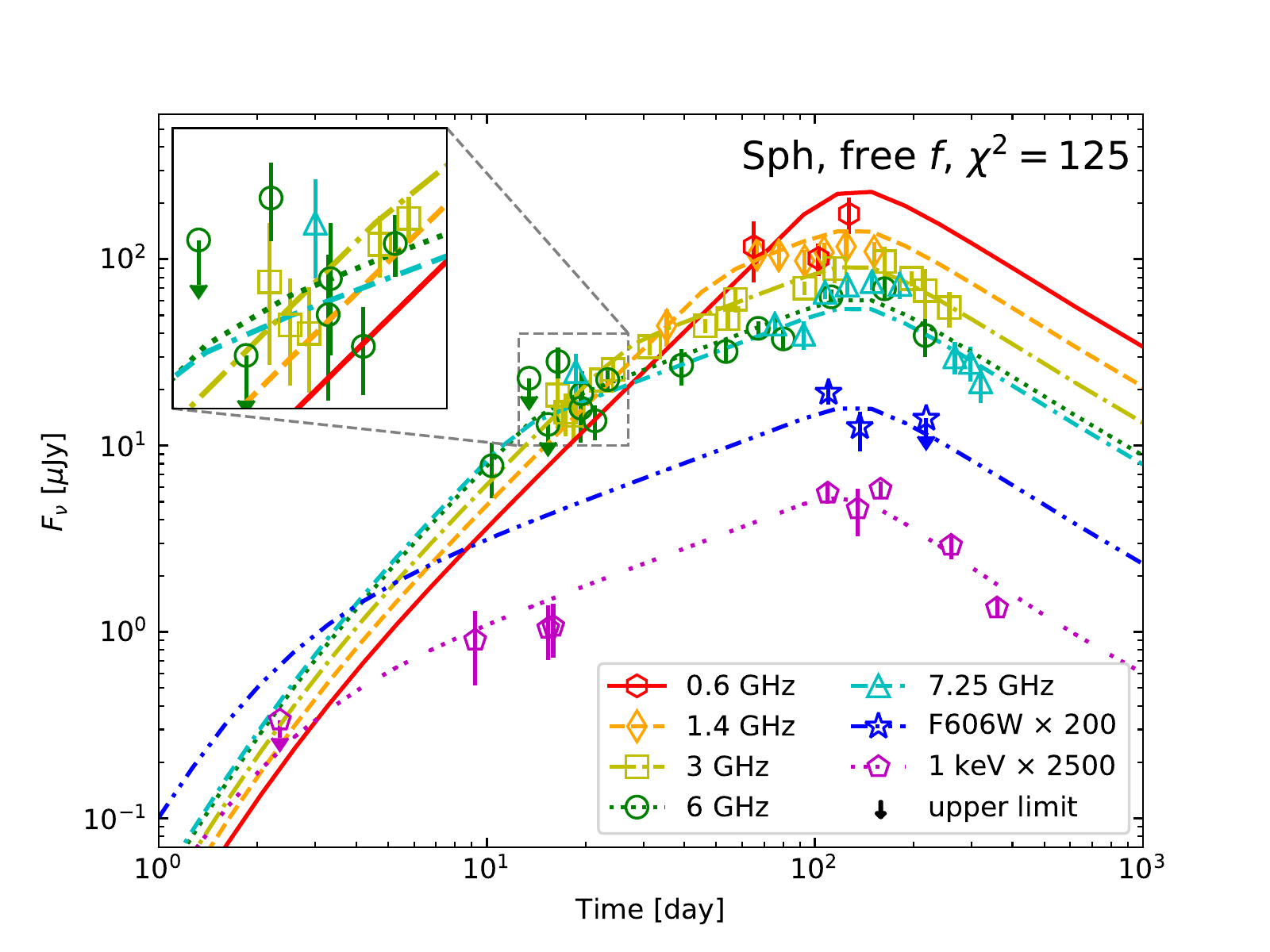}%\par
    \includegraphics[width=1\columnwidth]{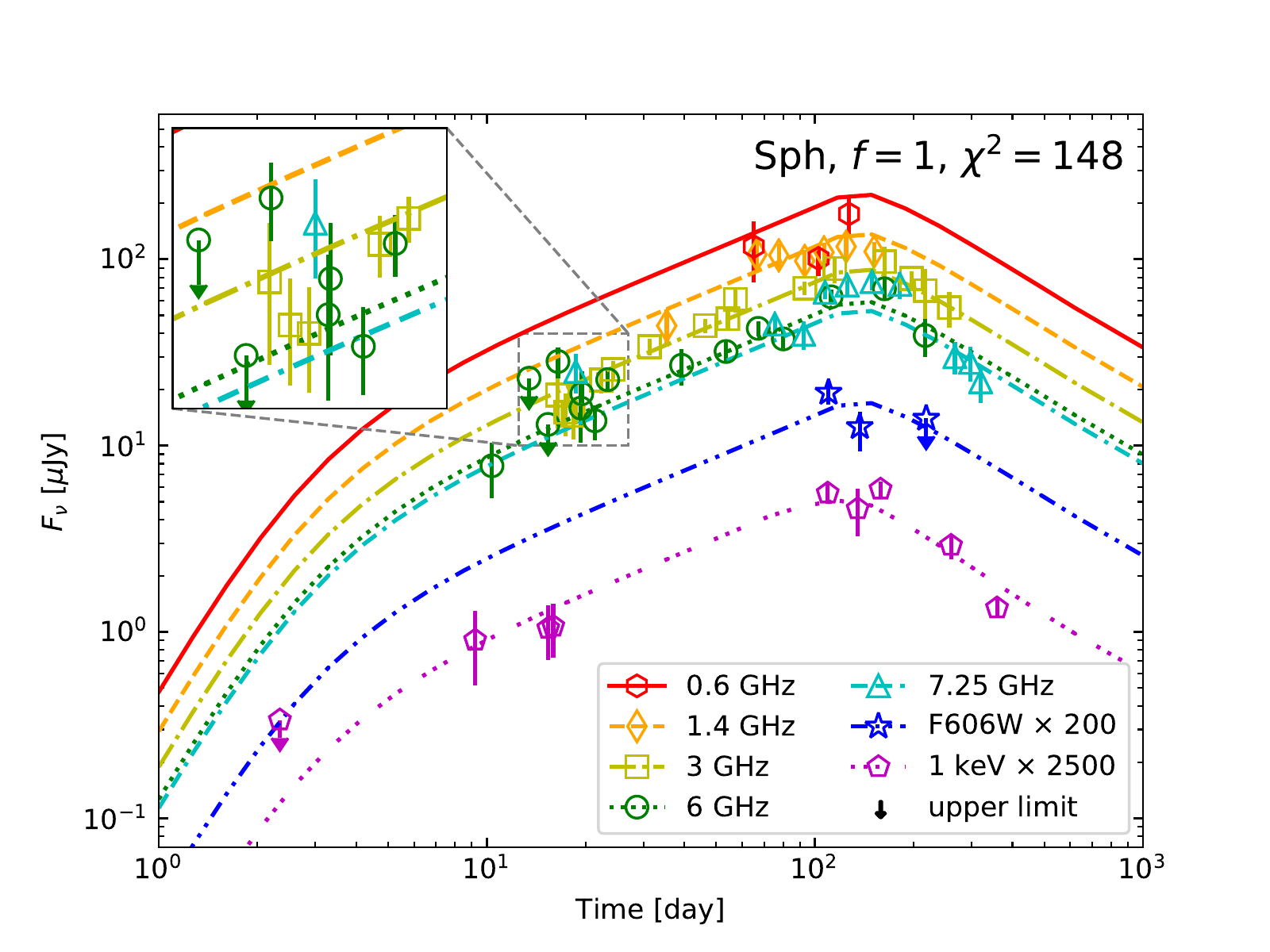}%\par
\end{multicols}
\caption{Radio, optical, and X-ray light curves of the best-fit (maximum likelihood) $f$ free jet model (top left), in comparison with the observed data points. Light curves of the same model but with $f=1$ and $\nu_{\rm obs} > \nu_m$ are shown in the top right panel. The bottom left panel shows light curves of the best-fit $f$ free spherical model, while the bottom right is the same but for the $f=1$ model. 
}
\label{fig:lc}
\end{center}
\end{figure*}

\section{Parameter estimation} \label{sec:mcmc}

We perform a Bayesian Markov-Chain Monte Carlo (MCMC) model fitting to the X-ray, optical and radio afterglow data of GW170817 up to $\sim 360$ days post-merger reported in \cite{2017ApJ...848L..21A, 2018ApJ...863L..18A, 2018A&A...613L...1D, 2018ApJ...858L..15D, 2017Sci...358.1579H, 2018NatAs...2..751L, 2018ApJ...856L..18M, 2018Natur.554..207M, 2018ApJ...867...57R, 2018MNRAS.tmpL..60T, 2018arXiv180806617V}. The total number of the data points is $N_{\rm data}=62$, $4$ of which are upper limits. Assuming a fixed distance of $D_L = 40$ Mpc, the synchrotron emission as a function of time and frequency is completely determined by the 9 parameters $(E_{\rm c,\,iso}, \Gamma_c, \theta_c, \theta_v, n, \epsilon_B, \epsilon_e, p, \eta_e)$ for the Gaussian structured jet model and $(E_{k,\,{\rm iso}}, k, u_{\rm max}, u_{\rm min}, n, \epsilon_B, \epsilon_e, p, \eta_e)$ for the stratified spherical outflow model. We further divided the modeling into two categories: one is an 8-parameter fit fixing the electron acceleration efficiency as $f=1$ like previous studies (the ``$f=1$ model'' hereafter), which is done by erasing the parameter $\eta_e$ by the constraint of $\eta_e = \epsilon_e(p-2)(p-1)$ as in eq.(\ref{f}). The other is the full 9-parameter fit with an additional parameter $\eta_e$, by which $f$ is allowed to vary freely (the ``$f$ free model'' hereafter).

We implement the public affine-invariant MCMC sampler package {\tt emcee} \citep{2013PASP..125..306F} to estimate the posterior distribution of parameters consistent with the observed data. We calculate $\chi^2$ goodness of fit and use the likelihood function ${\cal L} = \exp(-\chi^2/2)$ combined with uniform or log-uniform priors assigned for each parameter as the full probability function. The upper bound data are treated as zero flux with upper limits as the errors of corresponding confidence level. We initialize the MCMC walkers in a tiny Gaussian ball centered around a local maximum likelihood and generate $\sim 10^6$ samples to find an estimate of the posterior probability function. The parameter posterior distributions are summarized in Table \ref{tab:jet} and \ref{tab:sph}.

\section{Implications} \label{sec:imp}

\subsection{Difference by the $f$ free model}

\begin{figure}
  \begin{center}
    \includegraphics[width=\columnwidth]{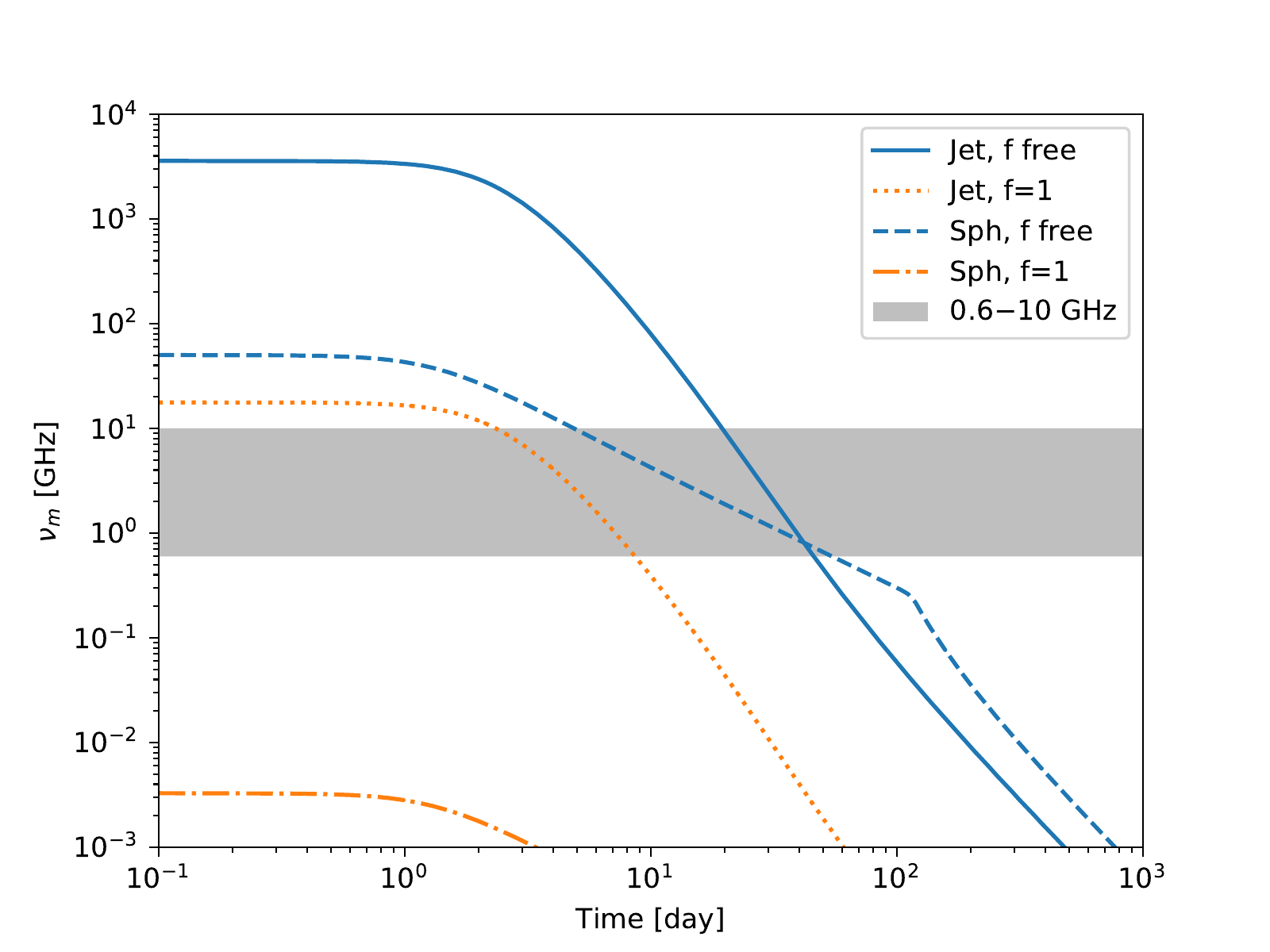}
  \end{center}
  \caption{Evolution of the characteristic synchrotron photon frequency $\nu_m$ corresponding to the lowest electron energy (in the observer frame) of the maximum likelihood solutions to the afterglow of GW170817, as shown in Fig. \ref{fig:lc}. The shaded region shows the observed radio bands of GW170817. With the conventional assumption $f=1$, $\nu_m$ is well below any radio band if the outflow is only mildly-relativistic.}
  \label{fig:nu_m}
\end{figure}

The best-fit (i.e., the maximum of the posterior probability density function) light curves by the $f$ free jet model are shown in Figure \ref{fig:lc} (top left panel). The physical parameters found with our new $f$ free model are in general similar to those by the conventional $f=1$ assumption. However, by allowing $\eta_e$ as a free parameter, the early radio data of GW170817 are in the low frequency tail ($\nu_{\rm obs} < \nu_m$ where $\nu_{\rm obs}$ is a frequency of observation), and as a results the light curve shape depends on the observed frequency. This is in sharp contrast to previous studies who found that all data are above $\nu_m$ and hence frequency-independent light curves. We also find that our best-fit $f=1$ jet model is in the low-frequency tail regime, but the nonthermal electron energy fraction is extremely high ($\epsilon_e \sim 1$), compared with the typical value $\epsilon_e \sim 0.1$ found in previous GRB afterglow studies \citep[e.g.][]{2015ApJ...815..102F, 2017MNRAS.472.3161B}. If we set a constraint of $\nu_{\rm obs} > \nu_m$ in addition to $f=1$, where $\nu_{\rm obs}$ is the observed bands (i.e. 0.6 GHz > $\nu_m$ during the observation period), we would find model parameters similar to those of previous studies (see the top right panel of Fig. \ref{fig:lc} and Table \ref{tab:jet}).

Similar trends about the $f$ parameter are found also for the spherical model, but in this case the best-fit $f=1$ model is in the $\nu_{\rm obs} > \nu_m$ regime with similar model parameters to previous studies. Therefore light curves of this model are shown in the bottom right panel of Fig. \ref{fig:lc}, in addition to those of the best-fit $f$ free model (bottom left). The $\chi^2$ is reduced from 140 ($f=1$) to 114 ($f$ free) for the jet model, while 148 to 125 for the spherical model. These are statistically significant difference for adding just one new model parameter. We note that using an exact synchrotron profile rather than the power-law approximation results in a similar reduction in the $\chi^2$ value. 

In the new $f$ free model, the ratio of the minimum electron energy to the post-shock proton energy, $\eta_e$, is strictly constrained to be around $\sim 0.1$ regardless of the outflow geometry, implying that the electrons are close to equipartition with the post-shock protons. This is consistent with the recent PIC simulations of relativistic shock \citep{2011ApJ...726...75S} and non-relativistic shock acceleration \citep{2015PhRvL.114h5003P}. The allowed region of $\epsilon_e$ includes the fiducial value $0.1$, but uncertainty is large and $\epsilon_e \sim 10^{-3}$ is also allowed within the statistical uncertainties. 
%see also \citealt{2018MNRAS.478.2281K}

The fact that the best-fit $f$ free model is found in the low frequency synchrotron tail regime indicates that $\nu_m$ is higher than the $f=1$ model. Indeed, $\nu_m$ from an on-axis outflow by the $f=1$ model is given as (in the relativistic limit)
\begin{equation}
  \nu_m(f=1) \simeq 2.2\,\,{\rm MHz\,} \epsilon_{e,\,-1}^2\epsilon_{B,\,-2}^{1/2}\,n_{-3}^{1/2}\,\Gamma_s^{4}
\end{equation}
with the convention of $Q_x = Q/10^x$ in cgs units, which is well below any monitoring band of GW170817 unless $\Gamma_s$ is sufficiently large. Nevertheless, by allowing $f$ free it becomes
\begin{equation} \label{nu_m}
  \nu_m(f\ {\rm free}) \simeq 11.6\,\,{\rm GHz\,} \eta_e^2\,\epsilon_{B,\,-2}^{1/2}\,n_{-3}^{1/2}\,\Gamma_s^{4} \ ,
\end{equation}
implying a possibility of $\nu_m$ higher than the GHz bands in the case of electron-proton equipartition ($\eta_e \sim 1$). The signature of the low-frequency tail regime would then appear in early radio data at low frequencies. We encourage observers to perform early low-frequency radio observations in the future events, to more clearly detect this signature. If detected, it would give important information about the electron-proton equipartition.

\subsection{Implications for the merger outflow}

\begin{figure}
  \begin{center}
    \includegraphics[width=\columnwidth]{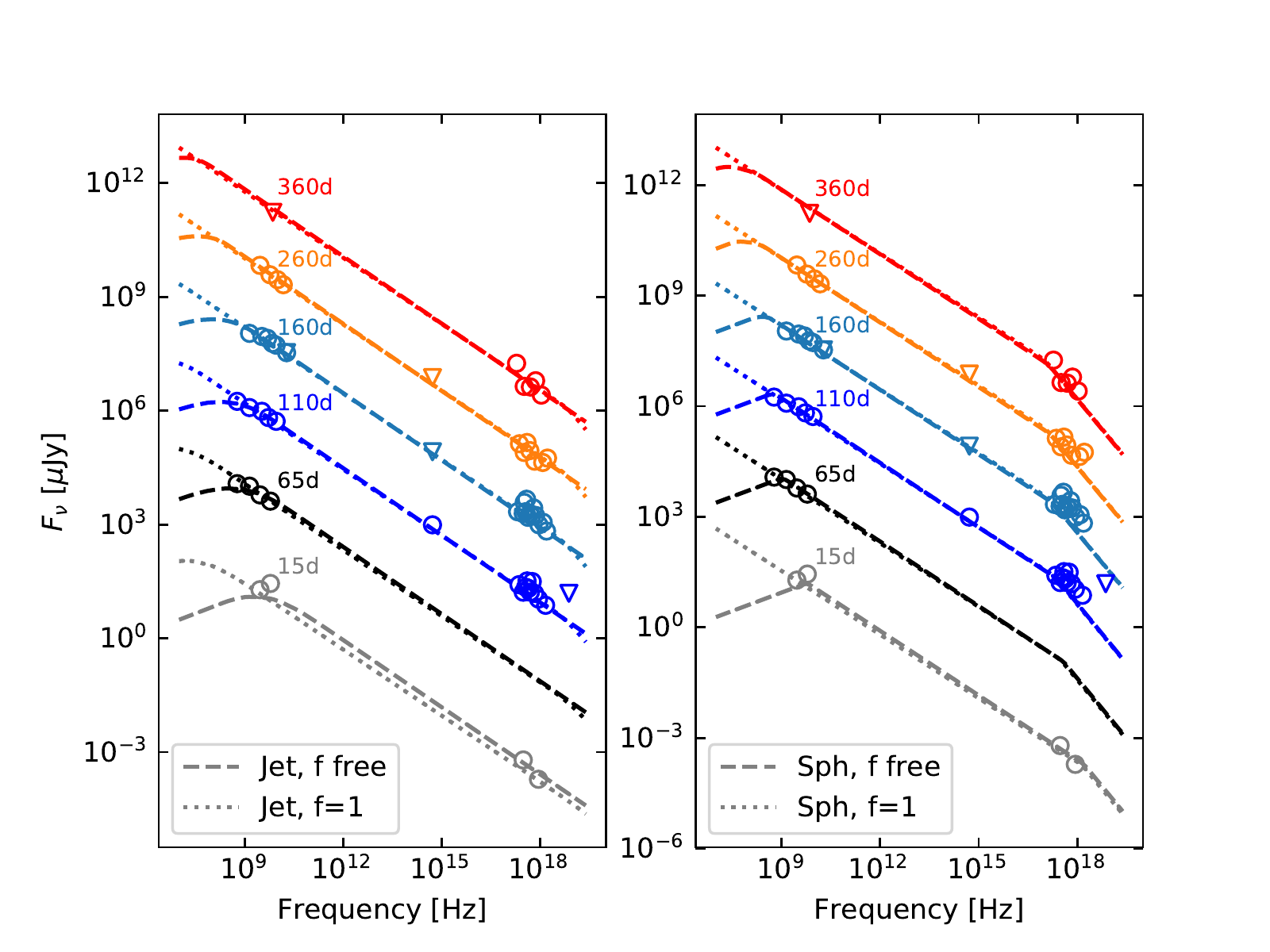}
  \end{center}
  \caption{Temporal evolution of the GW170817 afterglow spectral energy distribution at different time intervals, renormalized for legibility (upper limits are shown as inverted triangles), plotted with the maximum likelihood solutions shown in Fig.\ref{fig:lc}. }
  \label{fig:nu_m}
\end{figure}

The most prominent difference of the $f$ free model from the $f=1$ model is that the isotropic-equivalent energy to the jet direction and the total jet energy becomes larger by 1 -- 2 orders of magnitude (from $E_{c,\,{\rm iso}} = 10^{51.05}$ to $10^{52.38}$ erg and from $E_{\rm jet} = 10^{49.44}$ to $10^{50.74}$) in the jet model. However, both of these values are within the range of that found for short GRBs ($E_{\rm iso} \sim 10^{50}$ -- $10^{53}$ erg and the beaming corrected total energy $\sim 1.6^{+3.9}_{-1.3}\times 10^{50}$ erg \citep[e.g.][]{2015ApJ...815..102F}), and our preferred value with the $f$ free model is now at the high end, while that found by conventional modeling is at the low end.

The angle parameters of the jet model are not significantly changed by the $f$ free model. It favors a narrow jet with a half-opening angle $\theta_c \sim 0.08$ (= 4.6 deg) viewed off-axis at a viewing angle $\theta_v \sim 0.44$ (= 25.2 deg). Note that large angles are disfavored by the constraints of $\theta_v \lesssim 0.49$ set by LIGO \citep{2017PhRvL.119p1101A} and $\theta_v-\theta_c \lesssim 0.25$ imposed by the superluminal motion of the radio source near the time of peak flux \citep{2018Natur.561..355M}. We repeated our MCMC analysis with these constraints as a prior, the best-fit angles become $\theta_c \sim 0.05$ and $\theta_v \sim 0.29$, consistent with a previous estimation \citep{2018arXiv180610596H}. 

In the case of the spherical model, introducing the $f$ free model does not significantly change the outflow energy $E_{k,\,{\rm iso}}$ or the minimum velocity of the outflow. Compared with the kilonova ejecta (e.g. $E \sim 10^{51}$ erg and $\beta \sim$ 0.1--0.3, \citealt{2017ApJ...851L..21V}), the kinetic energy of the ejecta for the nonthermal afterglow is comparable, but the minimum velocity is faster ($u_{\rm min} = 10^{0.24}$ or $\beta_{\rm min} = 0.87$). This implies that the kilonova and the nonthermal afterglow of GW170817 must be powered by distinct ejecta components. The minimum cutoff velocity is not compatible with that of dynamical ejecta of BNS mergers ($\sim 0.4$ -- $0.5c$) suggested by numerical simulations \citep[e.g.][]{2017PhRvD..96h4060K}. A cocoon still remains as an alternative possibility, because most of its energy is expected to be carried by high velocity materials with $u \gtrsim 1$ \citep[e.g.][]{2018MNRAS.479..588G}.

\subsection{Ambient matter density}

As a common trend, the ambient density ($\log_{10} (n/{\rm cm}^{-3}) = -2.49^{+1.05}_{-1.08}$ and $-2.05^{+1.66}_{-1.71}$ for the jet and spherical models, respectively) found by the $f$ free model are about one order of magnitude higher than those found with the constraint of $f=1$. The high ambient density may seem in tension with the constraint $n \lesssim 0.04$ \pcm \ inferred from the HI mass observation of NGC 4993 \citep{2017Sci...358.1579H}. However, we note that a typical elliptical galaxy like NGC 4993 should be dominated by hot, ionized gas which is not constrained by HI observations. X-ray observations tell us that electron densities of hot gas in typical elliptical galaxies (e.g., NGC 1399 and 4472 whose absolute luminosities are similar to that of NGC 4993) are $\sim 0.1$ \pcm \ in the core region and $\sim 0.01$ \pcm \ at a half-light radius $r_e$ \citep{2003ARA&A..41..191M,2010MNRAS.404.1165C}. While there is no reported constraint on the hot gas in NGC 4993 \citep[see X-ray observations of the NGC 4993 in e.g.][]{2017ApJ...848L..22B, 2017ApJ...848L..25H}, based on the projected offset of the optical counterpart of GW170817 ($\sim 0.64r_e$, \citealt{2017ApJ...848L..22B}), the expected hot gas density should therefore be in the range of $10^{-2}$ -- $10^{-1}$ \pcm, consistent with the estimation by the $f$ free modeling. Rather, the low ambient density $\log_{10}(n/{\rm cm^{-3}}) = -3.71^{+0.61}_{-0.65}$ (jet) and $-3.68^{+1.21}_{-0.67}$ (spherical) found by the $f=1$ models are disfavored from this argument. 

\subsection{The cooling frequency}

\begin{figure}
  \begin{center}
    \includegraphics[width=\columnwidth]{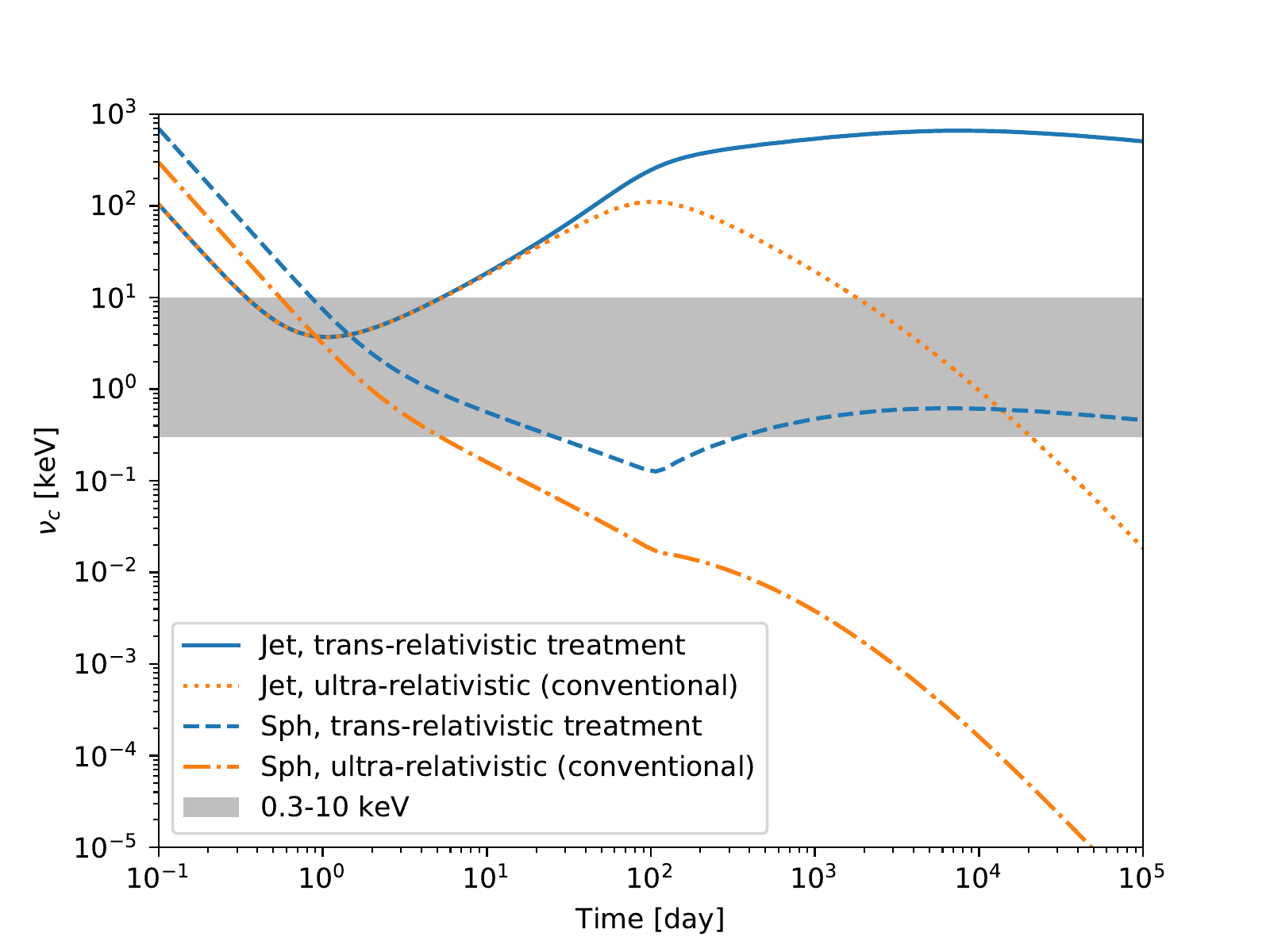}
  \end{center}
  \caption{Evolution of the cooling frequency $\nu_c$ (in the observer frame) with typical fitting parameters of the jet and spherical models to the afterglow of GW170187. (Here, the $f=1$ is shown but the difference from the $f$ free model is hardly visible in this plot.) The shaded region shows the observed X-ray bands for GW170817. A late-time monotonic decrease (orange) is predicted by the conventional GRB afterglow model which estimates magnetic field in a relativistic shock. However, if we correctly apply the trans-relativistic treatment for magnetic field as eq. (\ref{B}), the cooling break does not decrease any more (blue) once the shock decelerates to the mildly-relativistic regime at $\sim \mathcal{O}(100)$ days. 
}
  \label{fig:nu_c}
\end{figure}

The non-detection of cooling break in the X-ray spectrum of GW170817 up to 360 days after the merger \citep{2018ApJ...863L..18A} may disfavor the cocoon/dynamical ejecta origin modeled by the quasi-spherical outflow, which predicts a much earlier passage on a timescale no longer than a few months \citep{2018ApJ...867...95H}. However, we note that direct detection of a cooling break in GRB afterglow spectra are rare \citep[e.g.][]{2010ApJ...716L.135C}, and the cooling transition could be in fact smoother than expected by simple modelings if more realistic effects are taken into account, such as inclusion of evolutionary histories of comoving magnetic fields \citep{2014ApJ...780...82U}. Therefore, we cannot rule out the possibility of a much slower evolution of cooling break in the afterglow spectrum of GW170817.

Furthermore, we stress that the cooling frequency does not monotonically decrease in a mildly- or non-relativistic shock, in contrast to a ultra-relativistic shock usually considered in the conventional GRB afterglow theory. Time evolution of the cooling frequency in the trans-relativistic regime was not discussed in detail by previous studies for the nonthermal afterglow of GW170817. By eq.(\ref{B}), the comoving magnetic field strength should scale linearly with the shock velocity as $B' \propto \beta_s$ in the non-relativistic limit, and as a result the cooling frequency evolves as $\nu_c \propto \beta_s^{-3}t^{-2}$. As the shock enters the Sedov-Taylor phase, it becomes $\nu_c \propto t^{-1/5}$ and stops to decrease significantly (see Fig. \ref{fig:nu_c}). In this case, we are not supposed to see cooling passage in the near future, even if the cooling frequency already reached the X-ray regime ($\sim$ keV bands) as indicated by the spherical model fit. Therefore, passage time of the cooling frequency cannot be used as a diagnostic tool to discriminate between the jet and spherical models.

\section{Conclusion} \label{sec:con}

We propose a new formulation for the nonthermal electron energy distribution in a mildly relativistic shock driven by ejecta from a BNS merger, in which the number fraction $f$ of electrons injected into particle acceleration is freely variable. Then the minimum electron Lorentz factor $\gamma_m$ of the distribution and the total energy fraction carried by nonthermal electrons, $\epsilon_e$, become independent model parameters. This essentially adds another degree of freedom to the conventional modeling assuming that all electrons in the shock are accelerated ($f=1$) and hence $\epsilon_e$ is determined if $\gamma_m$ and other model parameters are fixed. In our model, $\gamma_m$ is determined by the degree of electron-ion equipartition, while $\epsilon_e$ is an independent parameter determined by the injection efficiency of electron acceleration.

In the framework of the two representative models (a Gaussian structured jet and a radially-stratified spherical outflow), we performed a MCMC model fitting to the X-ray/optical/radio afterglow data of GW170817 up to $\sim 360$ days post-merger. Previous studies with the conventional modeling found that the observed radio frequencies are always higher than $\nu_m$ (corresponding to $\gamma_m$), but we find that in the maximum likelihood solution of our new model the early radio fluxes are in the regime of the low-frequency tail ($\nu < \nu_m$). The new solution allows us to constrain $\eta_e$, the ratio of minimum electron energy to the post-shock proton energy, to be $\sim \mathcal{O}(10^{-1})$. This implies that there is a significant energy transfer from protons to electrons (otherwise we expect $\eta_e \sim 10^{-3}$), but still the full equipartition is not yet reached. It is also found that the best-fit of the electron fraction injected to particle acceleration is significantly smaller than the unity ($f \sim$ 0.01--0.1). These findings are consistent with previous PIC simulation of relativistic and non-relativistic shock \citep[e.g.][]{2011ApJ...726...75S, 2015PhRvL.114h5003P}.

While the posterior estimations of model parameters by the $f$ free model generally resemble those by the $f=1$ model, a notable difference is that in the isotropic-equivalent energy into the jet direction in the jet model is increased by at least one order of magnitude. The isotropic-equivalent energy inferred for GW170817 is still consistent with the range found for short GRBs, but now it is at the high end. Another notable difference is about one order of magnitude higher ambient density for the $f$ free model. The increased density may seem to be in tension with the constraint $n \lesssim 0.04$ \pcm \ inferred from the HI observation of the host galaxy \citep{2017Sci...358.1579H}, but we argue that the ambient density is likely dominated by hot X-ray emitting gas at the location of GW170817 in an elliptical galaxy, with a density ($\sim 10^{-2} - 10^{-1}$ \pcm) consistent with our model. 

We have shown that, by incorporating a trans-relativistic treatment of the comoving magnetic field strength in the shock front, the cooling frequency hardly evolves once the shock enters the mildly- or non-relativistic regime, as opposed to the standard GRB afterglow theory in the ultra-relativistic limit that predicts a monotonic decrease. The absence of the cooling break signature in the afterglow spectrum of GW170817 was utilized to argue against the stratified spherical outflow, but this result suggests that such an argument does not simply apply.

The decrease of $\chi^2$ by our new $f$ free model from the conventional $f=1$ model is larger than that expected by adding one new degree of freedom, implying that our new model is favored over the conventional model by the GW170817 data. However, the validity of our new model cannot be clearly seen as an early passage of $\nu_m$ in the radio bands, because of the paucity of early radio data ($\lesssim$ 10 days after merger). It would be difficult either to distinguish between the two models by the future evolution of GW170817. Therefore we encourage early, multi-band, and densely sampled radio observations of afterglows of future BNS merger events, which can be taken into consideration in radio follow-up strategy \citep[e.g.][]{2018ApJ...867..135C}. An unambiguous detection of $\nu_m$ passage would provide an essentially new constraint, giving important information for the degree of electron-ion equipartition and injection efficiency of particle acceleration.

\section*{Acknowledgements}

HL was supported by a MEXT scholarship. TT was supported by JSPS KAKENHI Grant Numbers JP15K05018, JP17H06362, and 18K03692. KK was supported by Grant-in-Aid for Scientific Research (16H02183, 17H06361, 18H01213) of JSPS and by a post-K computer project (Priority issue No. 9) of Japanese MEXT. 
Numerical simulations were performed on K computer at AICS (project numbers hp160211, hp170230, hp170313, hp180179), on Cray XC30 at CfCA of National Astronomical Observatory of Japan, Oakforest-PACS at Information Technology Center of the University of Tokyo, and on Cray XC40 at Yukawa Institute for Theoretical Physics, Kyoto University. 

\bibliographystyle{mnras}
%\bibliography{refs}
\input{refs.bbl}

\begin{figure*}
\begin{center}
\includegraphics[width=\textwidth]{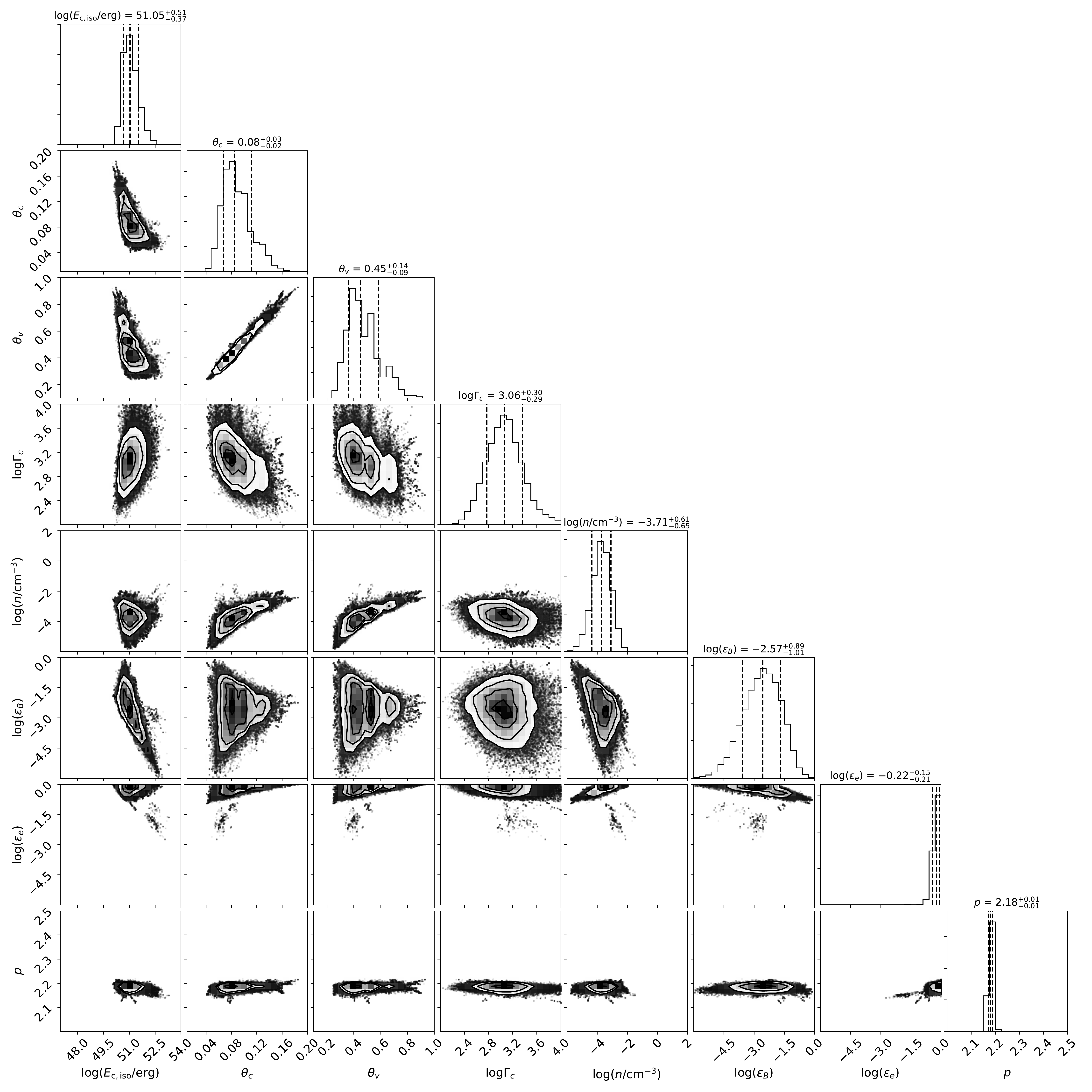}
\end{center}
\caption{Posterior probability distributions of parameters by Gaussian structured jet model with the conventional assumption $f=1$, visualized using the public plot tool {\tt corner} \citep{2016JOSS....1...24F}}. Contours of 0.5, 1, 1.5, 2-$\sigma$ are shown in the two-dimensional space of all possible combinations of two model parameters. The median values and their 1-$\sigma$ uncertainties are indicated in the one-dimensional distribution of each parameter.
\label{fig:corner_jet_f1}
\end{figure*}

\begin{figure*}
\begin{center}
\includegraphics[width=\textwidth]{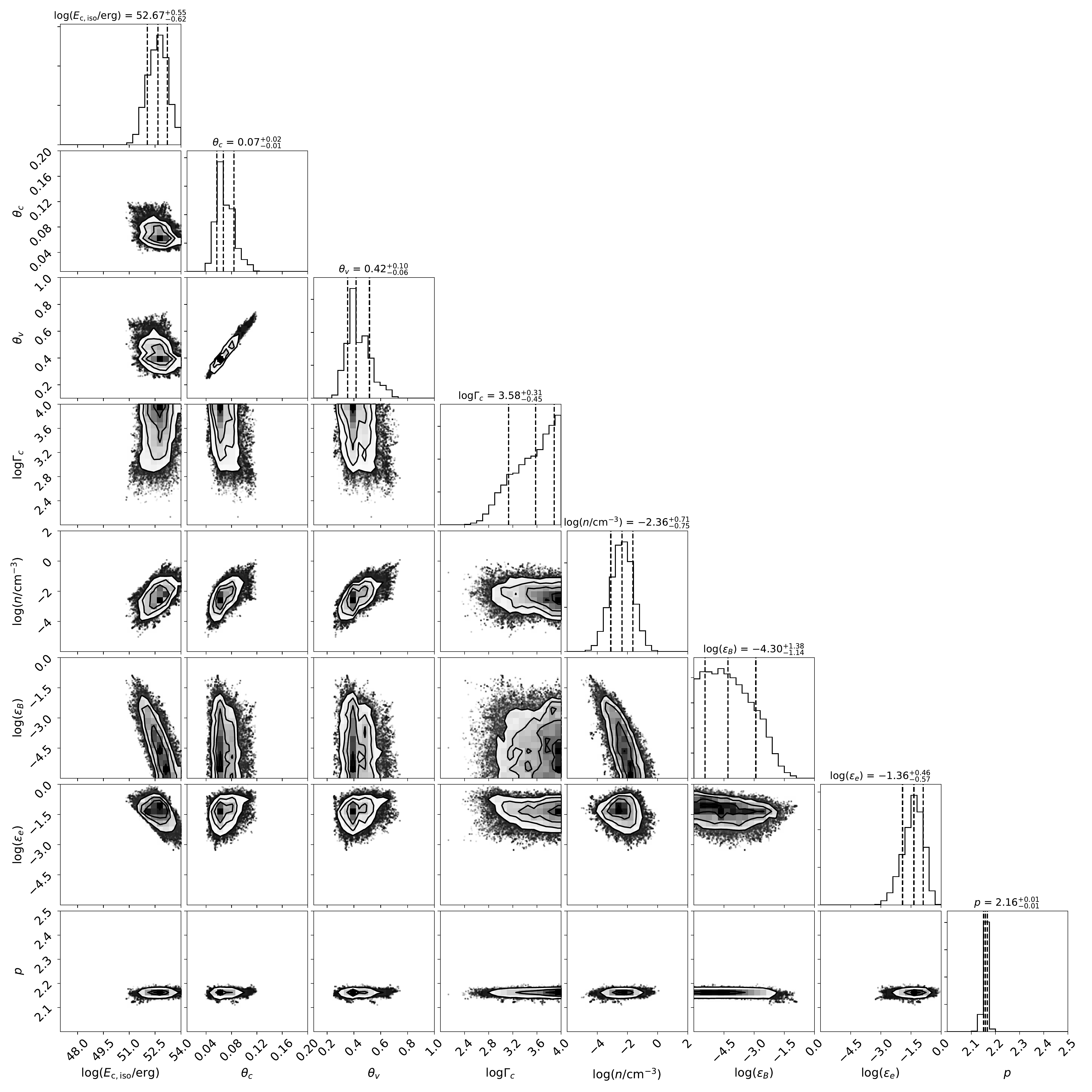}
\end{center}
\caption{The same as Figure \ref{fig:corner_jet_f1} by the $f=1$ Gaussian structured jet model, but with the additional constraint $\nu_{\rm obs} > \nu_m$, similar to previous studies. }
\label{fig:corner_jet_f1_nu}
\end{figure*}

\begin{figure*}
\begin{center}
\includegraphics[width=\textwidth]{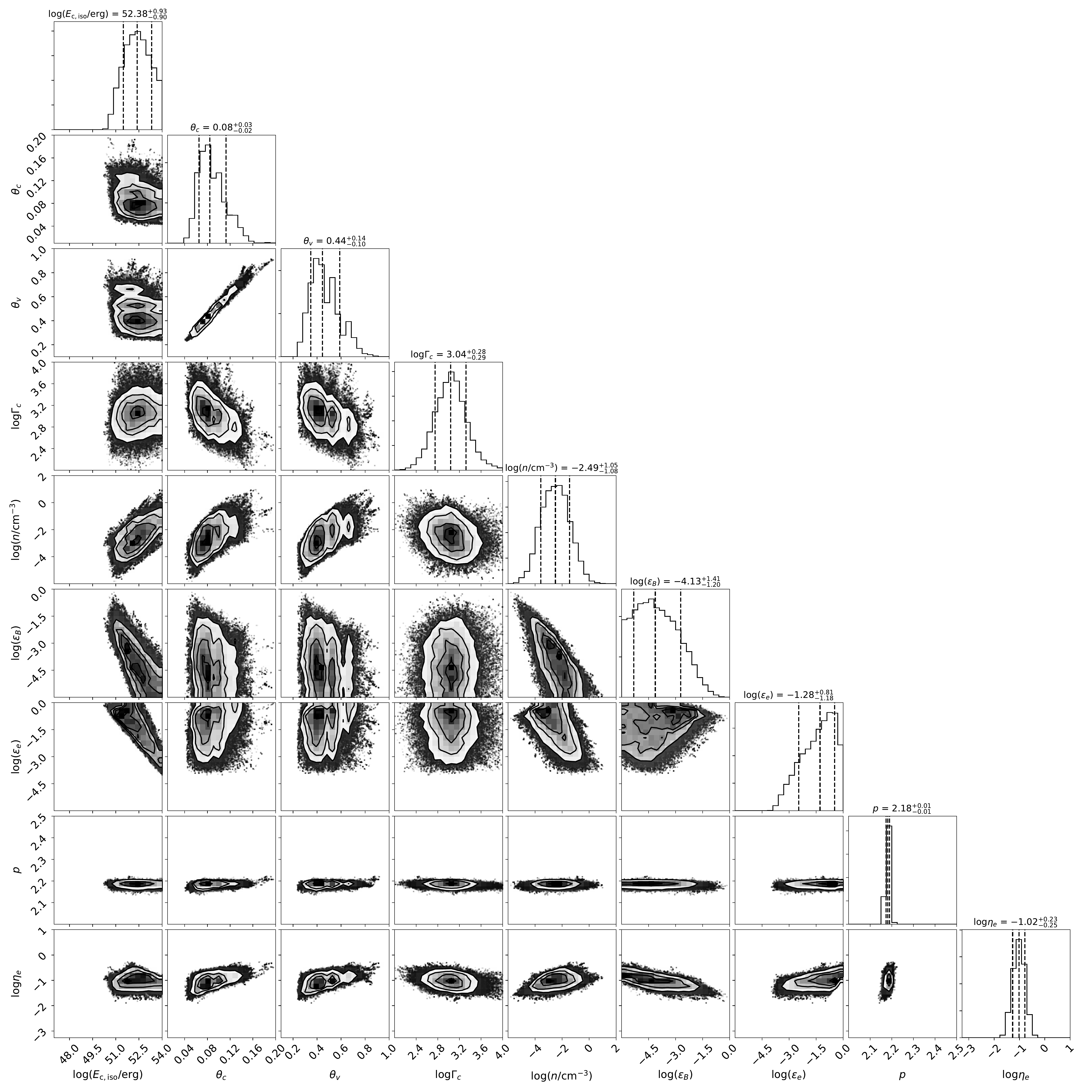}
\end{center}
\caption{The same as Figure \ref{fig:corner_jet_f1} by the $f$ free Gaussian structured jet model.}
\label{fig:corner_jet_ff}
\end{figure*}

\begin{figure*}
\begin{center}
\includegraphics[width=\textwidth]{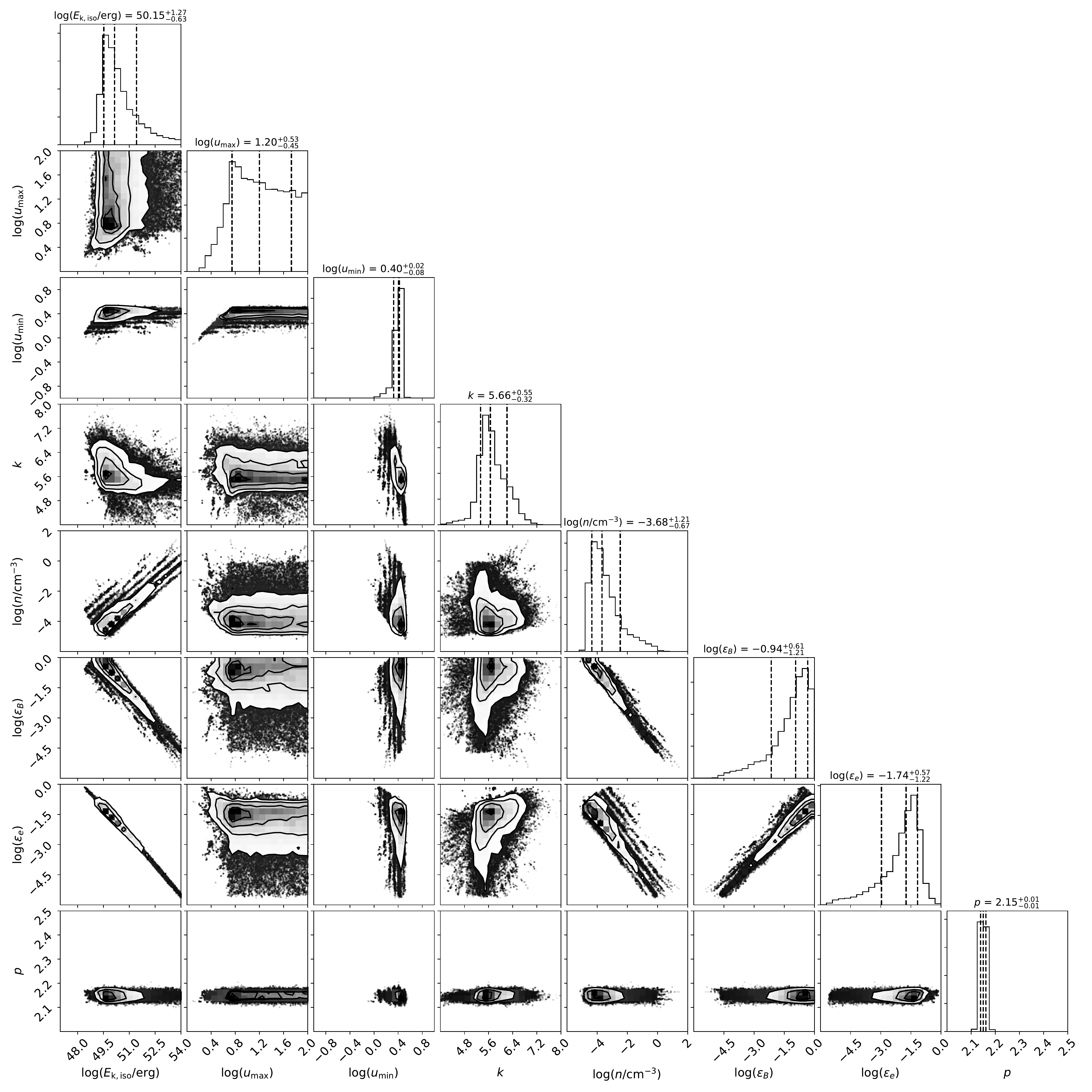}
\end{center}
\caption{The same as Figure \ref{fig:corner_jet_f1} but with stratified spherical outflow model ($f=1$). }
\label{fig:corner_sph_f1}
\end{figure*}

\begin{figure*}
\begin{center}
\includegraphics[width=\textwidth]{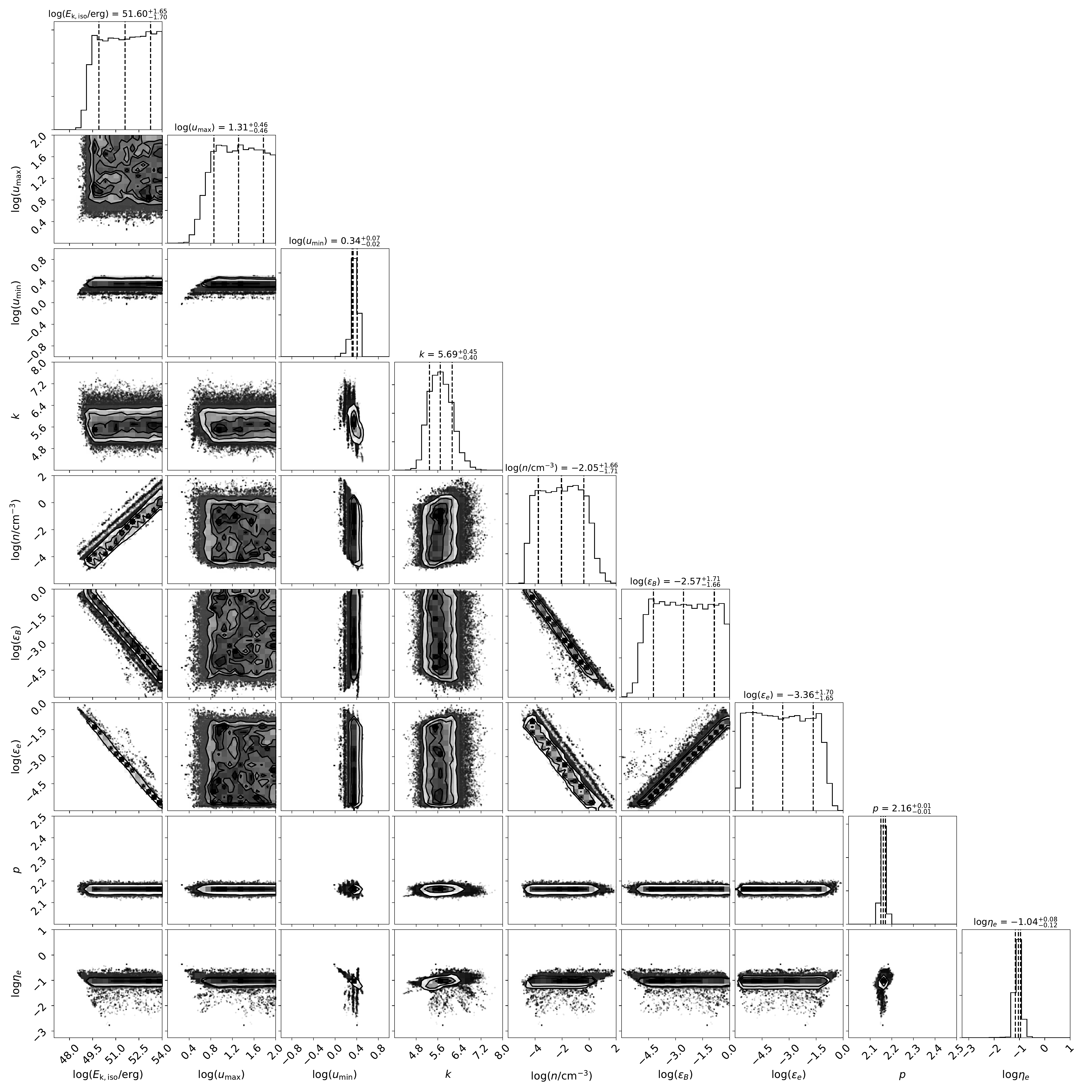}
\end{center}
\caption{The same as Figure \ref{fig:corner_jet_ff} but with stratified spherical outflow model ($f$ free). }
\label{fig:corner_sph_ff}
\end{figure*}

\bsp
\label{lastpage}
\end{document}